# Probabilistic Jamming/Eavesdropping Attacks to Confuse a Buffer-Aided Transmitter-Receiver Pair

Ahmed El Shafie†, Kamel Tourki*, Zhiguo Ding*, Naofal Al-Dhahir†

†Electrical Engineering Dept., University of Texas at Dallas, USA.
*Mathematical and Algorithmic Sciences Lab, France Research Center, Huawei Technologies Co. Ltd.
*School of Computing and Communications, Lancaster University, Lancaster, UK.

*Abstract*—We assume that a buffer-aided transmitter communicates with a receiving node in the presence of an attacker. We investigate the impact of a radio-frequency energy-harvesting attacker that probabilistically operates as a jammer or an eavesdropper. We show that even without the need for an external energy source, the attacker can still degrade the security of the legitimate system. We show that the random data arrival behavior at the transmitter and the channel randomness of the legitimate link can improve the system's security. Moreover, we design a jamming scheme for the attacker and investigate its impact on the secure throughput of the legitimate system. The attacker designs his power splitting parameter and jamming/eavesdropping probability based on the energy state of the attacker's battery to minimize the secure throughput of the legitimate system.

## I. INTRODUCTION

Radio-frequency (RF) energy-harvesting (EH) schemes enable charging of wireless nodes by exploiting energy of the ambient RF signals, where a wireless node converts the received RF transmissions into a direct current (DC). The two most promising schemes for RF energy harvesting are the time-switching scheme [1], where the receiver switches between data reception and energy harvesting, and the power-splitting scheme [1], where the receiver splits the signal into two streams of different powers for decoding information and harvesting energy separately to enable simultaneous information decoding and energy harvesting.

Communication secrecy in an information-theoretic sense was first investigated in the seminal work of Wyner [2] which is known as physical (PHY) layer security. The PHY security is measured by the secrecy rate which is determined by the difference between the rates of the legitimate link and the eavesdropping link.

Several works have investigated the impact of specific *active* malicious attacks such as denial-of-service (DoS) attacks [3], replay attacks [4], and data injection attacks [5]. The DoS attack, which blocks the communication between legitimate transceivers, has received great attention since it is the most accomplishable [3], [6], [7]. Recently, several researchers have investigated the linear quadratic Gaussian (LQG) control cost function problems under DoS attacks [3], [8]. In [3], the authors proposed a semi-definite programming-based solution to design a feedback controller that minimizes a cost function subject to energy constraints. The authors of [8] investigated the zero-sum game between a controller and a jammer.

The authors of [9] investigated the optimal jamming attack strategies by optimizing the probability of jamming and transmission range. The author of [10] investigated the impact of active jamming attacks on underwater wireless sensor networks. In [11], the authors designed an optimal attack schedule to maximize the attacking effect on the wireless networked control system (WNCS). The DoS attacker has to make the attack decision, i.e., when to attack and which channel to be chosen. In contrast to all the above-mentioned works, we assume that the attacker can sometimes remain silent to intercept the communication between the legitimate nodes. In addition, we assume that the eavesdropper exploits the energy of the RF signals from the legitimate system to power his battery.

In this letter, we design an opportunistic attack scheme to maximize the attacking impact on the legitimate system. The attacker is assumed to be an RF-EH node with infinite size battery. Our main message is that an attacker with no external energy source can still significantly degrade the system security. We study the impact of the channel and data arrival randomness at the legitimate transmitter on the attacking mode. In our investigated scenario, the attacker has to make the jamming or eavesdropping attack decision, i.e., jamming the channel to cause DoS or eavesdropping on the transmissions. The main contributions of this letter, which distinguish it from the related literature, are summarized as follows

- We formulate a joint DoS/eavesdropping attack problem and develop the optimal attack scheme that minimizes the secure throughput of the legitimate system. In our analysis, we consider the decoding and processing energies at the eavesdropper's receiver as well as the eavesdropper's battery dynamics. Furthermore, we investigate the impact of the data buffer at the legitimate source node on the system security.
- We design an access scheme at the legitimate system to mitigate the attacks, which is important to reduce the energy transfer rate to the attacker and helpful in draining his battery.
- We derive closed-from expressions for the mean service rate of the transmitter's data queue, legitimate system secure throughput, and the energy arrival and departure rates at the eavesdropping node.

## II. SYSTEM MODEL AND PROPOSED SCHEME

We consider the following wireless transmission scenario. A buffered source node (Alice) wishes to transmit her data to her receiving node (Bob) privately from a malicious node (Eve). Time is partitioned into time slots where a time slot equals the channel coherence time duration. Eve is assumed to be an RF-EH node as in [12]. Hence, she converts the ambient RF transmissions into energy. Moreover, we consider the general case where a constant energy source is also available at Eve.

This paper was made possible by NPRP grant number 8-627-2-260 from the Qatar National Research Fund (a member of Qatar Foundation). The statements made herein are solely the responsibility of the authors.

That is, an amount of $E_{\text{const}}$ energy units is supplied to Eve per time slot. She needs at least $E_{\text{d}} = P_{\text{d}}T$ energy units to decode Alice's data during a time slot of $T$ seconds where $P_{\text{d}}$ denotes the amount of power consumed by filters, frequency synthesizer, etc. If Eve's battery has energy lower than $E_{\text{d}}$, she either harvests energy without power splitting or she jams the channel between Alice and Bob. The jamming is beneficial to Eve since she will force Alice to retransmit the data packets and block Alice's buffer, which can be used to further increase the energy collected at Eve's receiver. If Eve's battery has energy higher than $E_{\text{d}}$, Eve will either receive data and harvest energy or jam the Alice-Bob channel. We denote the battery at Eve as $\mathcal{B}_{\text{E}}$. Alice transmits her data with average power $P_{\text{A}}$ Watts and Eve transmits her jamming signal with average power $P_{\text{J}}$ Watts.

Assume that the data buffer at Alice is denoted by $Q_{\text{A}}$ where each packet has a fixed size of $b$ bits. To capture the random nature of the data arrivals at the nodes, we assume that the data arrivals at Alice are Bernoulli random variables with mean $\lambda_{\text{A}}$ packets/slot. The Bernoulli arrival model has been commonly used for modeling the arrival process at source nodes to capture the data traffic burstiness [13].[1] To transmit a data packet during a time slot with duration $T$ seconds and a channel bandwidth of $W$ Hz, the target secrecy rate is $\mathcal{R} = b/(WT)$ bits/sec/Hz. Since we do not assume the availability of Eve's instantaneous channel state information (CSI) at Alice, Alice can only choose a target secrecy rate to design the random binning scheme which achieves the instantaneous secrecy rate. The codebook rate of the random binning scheme, denoted by $\mathcal{R}_{\text{A}}$, is set to the instantaneous rate of the Alice-Bob link. If the instantaneous rate of the Alice-Bob link is higher than the target secrecy rate $\mathcal{R}$, Alice will transmit the data packet and there is no connection outage. If the rate of the Alice-Bob link is lower than $\mathcal{R}$, Alice will remain silent since the transmission will not be decoded at Bob and Eve will take advantage of these transmissions to power her receiver. At the end of a time slot, the legitimate receiver (Bob) sends a feedback signal to acknowledge the correct/fail decoding of the data packet [13].

The thermal noise is modeled as additive white Gaussian noise (AWGN) with zero mean and power $\kappa$ Watts. We assume the flat-fading channel model where each channel coefficient remains unchanged for the time slot (coherence time) duration, but changes from one slot duration to another independently. Since the channels are randomly generated, Alice does not know the CSI of Eve when Eve operates as an eavesdropper (i.e. during the time slots where Eve decides to eavesdrop, she will remain completely silent, hence, Alice and Bob will not be able to estimate Eve's channel). Moreover, Eve does not know the channel between Alice and Bob since we assume that channel estimation is realized using pilots transmitted from one side and the channel is estimated from the other side. Since the channel between Alice and Bob is random and can be in outage, Alice will transmit only when her channel is not in outage. Hence, Alice saves her power and this strategy might hurt Eve since Eve selects either the *data eavesdropping* decision or the *jamming* decision and, hence, she will lose energy $E_{\text{d}}$ or $E_{\text{J}} \geq E_{\text{d}}$, respectively. Similarly, the data traffic burstiness and the data buffer state at Alice can lead to energy losses at Eve. This shows that the channel and data arrival randomness can be useful to secure the system. Based on this observation, we suggest the use of a random access scheme at Alice when the channel is not in outage. This helps in draining the battery of Eve and, hence, she will not be able to decode Alice's data or jam Bob.

[1] Changing the arrival model will only change the probability of the queue being empty/nonempty. However, the rate expressions and the analysis are unchanged.

Eve decodes the data at energy cost $E_{\text{d}}$. If she does not have this amount of energy in her battery, she will not be able to decode the data transmitted by Alice. Hence, Eve harvests energy. If Eve's battery level is higher than $E_{\text{d}}$ but lower than $E_{\text{J}}$, she will harvest and decode data at the same time using the power splitting scheme. We define $\rho$ as the amount of power fraction used for data decoding. If the energy state at Eve's battery is higher than $E_{\text{J}}$, Eve can either harvest energy and decode data with probability $1 - \alpha_{\text{E}}$ or jam Alice's transmission with probability $0 \leq \alpha_{\text{E}} \leq 1$.

### III. SYSTEM ANALYSIS

The mean service rate of Alice's queue in packets/slot is

$$\mu_{\text{A}} = (\alpha_{\text{A}}(1-\alpha_{\text{E}})\Pr\{\mathcal{B}_{\text{E}} \geq E_{\text{J}}\} + \alpha_{\text{A}}\Pr\{\mathcal{B}_{\text{E}} < E_{\text{J}}\})\mathbb{P}_1 \quad (1)$$

where $\gamma_{\text{A}} = P_{\text{A}}/\kappa$, $\mathbb{P}_1 = \Pr\{\log_2\left(1 + \gamma_{\text{A}}|h_{\text{A-B}}|^2\right) \geq \mathcal{R}\}$ which equals $\exp\left(-\frac{2^{\mathcal{R}}-1}{\sigma_{\text{A-B}}^2 \gamma_{\text{A}}}\right)$ for Rayleigh channel model with $\sigma_{\text{A-B}}^2$ denoting the Alice-Bob channel variance, $\{\log_2\left(1 + \gamma_{\text{A}}|h_{\text{A-B}}|^2\right) \geq \mathcal{R}\}$ is the event that the Alice-Bob link can support the target secrecy rate $\mathcal{R}$, $|h_{\text{A-B}}|^2\gamma_{\text{A}}$ is the received signal-to-noise ratio (SNR) at Bob with $h_{\text{A-B}}$ denoting the channel coefficient between Alice and Bob, $\alpha_{\text{A}}$ is the probability that Alice decides to access the channel, and $\alpha_{\text{E}}$ is the probability that Eve decides to jam. Note that if there is interference, and since Alice designs the codebook rate of the random binning scheme based on the instantaneous rate of the Alice-Bob link when there is no jamming (since the attack type is unknown a priori at Alice), there is a connection outage and, hence, the throughput and queue service rate is zero. If $\{\mathcal{B}_{\text{E}} < E_{\text{J}}\}$, Eve will either harvest and decode when $\{E_{\text{d}} \leq \mathcal{B}_{\text{E}} < E_{\text{J}}\}$ or harvest only when $\{\mathcal{B}_{\text{E}} < E_{\text{d}}\}$.

Alice is active in a time slot if her queue is nonempty, she decides to access the channel, and the channel of the Alice-Bob link is not in outage. We denote Alice's activity process by $\mathbb{I}_{\text{A}}$, which equals to $1$ when Alice is active and zero otherwise. We also define $\mathbb{I}_{\text{E}}$ as the activity indicator of Eve. Based on the description given above, the energy harvested at Eve is given by

$$\begin{aligned}E_{\text{H}} = {}& \eta|h_{\text{A-E}}|^2 P_{\text{A}} T \mathbb{I}_{\text{A}} 1(\mathcal{B}_{\text{E}} < E_{\text{d}}) \\ & + \eta(1-\rho)|h_{\text{A-E}}|^2 P_{\text{A}} T \mathbb{I}_{\text{A}} 1(E_{\text{d}} \leq \mathcal{B}_{\text{E}} < E_{\text{J}}) \\ & + \eta(1-\rho)|h_{\text{A-E}}|^2 P_{\text{A}} T \mathbb{I}_{\text{A}} 1(\mathcal{B}_{\text{E}} \geq E_{\text{J}})(1-\mathbb{I}_{\text{E}})\end{aligned} \quad (2)$$

where $0 \leq \eta \leq 1$ is the RF-to-DC conversion efficiency and $1(\cdot) = 1$ when the enclosed event is true.

When Eve decides to decode, she loses $E_{\text{d}}$ energy units. Furthermore, when she decides to jam, she loses $E_{\text{J}}$ units. The energy depletion at Eve is given by

$$E_{\text{out}} = \mathbb{I}_{\text{A}}\left(E_{\text{d}}1(E_{\text{d}} \leq \mathcal{B}_{\text{E}} < E_{\text{J}}) + \left(E_{\text{d}}\overline{\mathbb{I}_{\text{E}}} + E_{\text{J}}\mathbb{I}_{\text{E}}\right)1(\mathcal{B}_{\text{E}} \geq E_{\text{J}})\right) \quad (3)$$

where $\overline{\mathbb{I}_\text{E}} = 1 - \mathbb{I}_\text{E}$ and $E_\text{J} = P_\text{J} T$.

Eve's battery evolves as follows

$$\mathcal{B}(t+1) = \mathcal{B}(t) - E_\text{out} + E_\text{H} + E_\text{const} \quad (4)$$

where $\mathcal{B}(t+1)$ denotes the Eve's battery state at time slot $t+1$, and $E_\text{H} + E_\text{const}$ is the total energy collected at Eve in a given time slot.

The secure throughput in packets/slot is given by

$$\begin{aligned}\mu^\text{sec} = \alpha_\text{A} &\bigg( \Pr\{\mathcal{B}_\text{E} \geq E_\text{J}, Q_\text{A} > 0\}(1-\alpha_\text{E}) \\ &+ \Pr\{\mathcal{B}_\text{E} < E_\text{J}, Q_\text{A} > 0\} \bigg) \Pr\left\{R_\text{sec}^\text{nojam} \geq \mathcal{R}\right\} \\ &+ \alpha_\text{A} \Pr\{\mathcal{B}_\text{E} < E_\text{d}, Q_\text{A} > 0\} \Pr\left\{\log_2\left(1+\gamma_\text{A}|h_\text{A-E}|^2\right) \geq \mathcal{R}\right\} \\ &+ \alpha_\text{A} \alpha_\text{E} \mathbb{P}_2 \Pr\{\mathcal{B}_\text{E} \geq E_\text{J}, Q_\text{A} > 0\}\end{aligned} \quad (5)$$

where $\mathbb{P}_2 = \Pr\left\{\log_2\left(1 + \frac{|h_\text{A-B}|^2 \gamma_\text{A}}{1+|h_\text{E-B}|^2 \gamma_\text{E}}\right) \geq \mathcal{R}\right\}$, $\gamma_\text{E} = P_\text{J}/\kappa$, and

$$R_\text{sec}^\text{nojam} = \left[\log_2\left(1+\gamma_\text{A}|h_\text{A-B}|^2\right) - \log_2\left(1+\rho\gamma_\text{A}|h_\text{A-E}|^2\right)\right]^+ \quad (6)$$

The second term in (6), which is Eve's rate, is a function of $\rho$ since Eve uses a power splitter. When there is jamming, Eve will not eavesdrop. Hence, the security is achieved when the direct link (i.e. Alice-Bob link) is not in a connection outage. The closed-form expressions for the outage and secrecy outage probabilities under Rayleigh fading channel model can be easily derived and are omitted here due to space limitations.

Since Alice does not know the parameter $\rho$, we set the access probability of Alice such that Alice's queue is stable under a lower bound on mean service rate of Alice's queue. If the queue is stable using the lower bound on the service rate, then it will be stable for the actual operation. The queue stability is a widely-used quality-of-service (QoS) measure for buffered-aided nodes and is important here to decrease the access times of Alice and, hence, hurt Eve by reducing the energy collected at Eve's receiver. Let us first obtain a lower bound on $\mu_\text{A}$ in (1) as follows. Since $\Pr\{\mathcal{B}_\text{E} \geq E_\text{J}\} \leq 1$,

$$\mu_\text{A} \geq \alpha_\text{A}(1-\alpha_\text{E})\exp\left(-\frac{2^\mathcal{R}-1}{\sigma_\text{A-B}^2 \gamma_\text{A}}\right) \quad (7)$$

Using Loynes theorem, Alice's queue is stable when $\lambda_\text{A} \leq \alpha_\text{A}(1-\alpha_\text{E})\exp\left(-\frac{2^\mathcal{R}-1}{\sigma_\text{A-B}^2 \gamma_\text{A}}\right)$.

When Eve decides to harvest and decode, she can further enhance her performance by using a set of the collected data after only $\tau \leq T$ seconds to decide whether or not Alice is active. This will save Eve's energy since only a portion of the decoding and processing energy will be used. More specifically, assuming a processing power $P_\text{d}$, if the eavesdropper decides to harvest for the whole transmission time, the used energy in decoding is $P_\text{d} T$. However, if the eavesdropper uses only $\tau$ seconds to decide, the used energy is $P_\text{d} \tau \leq P_\text{d} T$ with equality when $\tau = T$. Assuming that Eve uses an energy detector, the false alarm and miss-detection probabilities are given by $P_\text{FA}^\tau$ and $P_\text{MD}^\tau$, respectively. If Eve miss detects the activity of Alice, i.e., considers Alice inactive while she is active, Eve will stop decoding and harvesting.

Hence, she will lose both energy and rate. If there is a false alarm, Eve will consider Alice to be active while she is not active. Hence, Eve will waste her decoding energy without any gain. On the other hand, if there is a correct detection of Alice's activity, Eve will not waste time in decoding noisy signals when Alice is inactive. Increasing $\tau$ will decrease both miss-detection and false-alarm probabilities, which helps Eve. However, it will also increase the amount of energy used at Eve in processing before deciding whether or not Alice is active. The best choice for $\tau$ is challenging and Eve can optimize it, in addition to the jamming/harvesting probability and power splitting factor, to minimize the secure throughput of Alice. The secure throughput is given in (8) at the top of next page with $\mathbb{P}_3 = \Pr\left\{\log_2\left(1+\gamma_\text{A}|h_\text{A-E}|^2\right) \geq \mathcal{R}\right\}$.

Updating (2), the energy harvested at Eve under channel sensing is given by

$$\begin{aligned}E_\text{H} = \eta|h_\text{A-E}|^2 P_\text{A} T \mathbb{I}_\text{A} &\bigg( 1(\mathcal{B}_\text{E} < E_\text{d}) \\ &+ (1-\rho)(\tfrac{\tau}{T}\mathbb{I}_\text{MD} + \mathbb{I}_\text{D})1(E_\text{d} \leq \mathcal{B}_\text{E} < E_\text{J}) \\ &+ (1-\rho)1(\mathcal{B}_\text{E} \geq E_\text{J})(1-\mathbb{I}_\text{E})(\tfrac{\tau}{T}\mathbb{I}_\text{MD} + \mathbb{I}_\text{D}) \bigg)\end{aligned} \quad (9)$$

where $\mathbb{I}_\text{MD} = 1 - \mathbb{I}_\text{D} = 1$ when Eve miss detects the activity of Alice. The energy depletion process at Eve is given by

$$\begin{aligned}E_\text{out} = &E_\text{J} \mathbb{I}_\text{E} 1(\mathcal{B}_\text{E} \geq E_\text{J}) \\ &+ E_\text{d}\left(\mathbb{I}_\text{A}\left(\mathbb{I}_\text{MD}\tfrac{\tau}{T} + \mathbb{I}_\text{D}\right) + \overline{\mathbb{I}_\text{A}}\left(\overline{\mathbb{I}_\text{FA}}\tfrac{\tau}{T} + \mathbb{I}_\text{FA}\right)\right) \\ &\times \left(1(E_\text{d} \leq \mathcal{B}_\text{E} < E_\text{J}) + \overline{\mathbb{I}_\text{E}} 1(\mathcal{B}_\text{E} \geq E_\text{J})\right)\end{aligned} \quad (10)$$

where $\mathbb{I}_\text{FA} = 1$ when Eve's sensor generates a false alarm. For a targeted false alarm probability, $P_\text{FA}^\tau = P_\text{FA}$, the probability of miss detection under Rayleigh fading channel model is given by

$$\begin{aligned}P_\text{MD}^\tau = 1 - &\frac{1}{\tilde{\gamma}_\text{A}}\exp\left(\frac{1}{\tilde{\gamma}_\text{A}}\right) \\ &\times \int_1^\infty \mathcal{Q}\left(\sqrt{f_\text{s}\tau}\bigg(\frac{\frac{Q^{-1}(P_\text{FA})}{\sqrt{f_\text{s}\tau}}+1}{\mathcal{Z}}-1\bigg)\right)\exp\left(-\frac{\mathcal{Z}}{\tilde{\gamma}_\text{A}}\right)d\mathcal{Z}\end{aligned} \quad (11)$$

where $f_\text{s} = 1/W$ is the sampling frequency, $\sigma_\text{A-E}^2$ denotes the Alice-Eve channel variance, $\tilde{\gamma}_\text{A} = \sigma_\text{A-E}^2 \gamma_\text{A}$, and $Q^{-1}(.)$ is the inverse of Q-function.

To minimize the secure throughput, Eve solves the following constrained optimization problem:

$$\min_{\alpha_\text{E}, \rho} : \mu^\text{sec}, \text{ s.t. } 0 \leq \rho, \alpha_\text{E} \leq 1 \quad (12)$$

The optimization problem is nonconvex due to the nonconvexity of the objective function. Eve can solve (12) **OFFLINE** using the interior-point methods, or a 2-dimensional grid-based search over the optimization variables. If Eve uses the 2-dimensional grid-based search, she will divide the range of each optimization variable into $M$ points, and then compute the objective function and the constraints for $M^2$ times since she optimizes over two optimization variables. The solution obtained from such grid-based search will maximize the objective function and satisfy the constraints. These computations

$$\begin{aligned}\mu^{\text{sec}} =& \alpha_{\text{A}}\left(\Pr\{\mathcal{B}_{\text{E}}\geq E_{\text{J}},Q_{\text{A}}>0\}(1-P_{\text{MD}}^\tau)(1-\alpha_{\text{E}}) + (1-P_{\text{MD}}^\tau)\Pr\{E_{\text{d}}\leq\mathcal{B}_{\text{E}}<E_{\text{J}},Q_{\text{A}}>0\}\right)\Pr\left\{R_{\text{sec}}^{\text{nojam}}\geq\mathcal{R}\right\} \\ &+ \alpha_{\text{A}}\left(\Pr\{\mathcal{B}_{\text{E}}\geq E_{\text{J}},Q_{\text{A}}>0\}P_{\text{MD}}^\tau(1-\alpha_{\text{E}}) + P_{\text{MD}}^\tau\Pr\{E_{\text{d}}\leq\mathcal{B}_{\text{E}}<E_{\text{J}},Q_{\text{A}}>0\} + \Pr\{\mathcal{B}_{\text{E}}<E_{\text{d}},Q_{\text{A}}>0\}\right)\mathbb{P}_3 \\ &+ \alpha_{\text{A}}\alpha_{\text{E}}\Pr\left\{\log_2\left(1+\frac{|h_{\text{A}-\text{B}}|^2\gamma_{\text{A}}}{1+|h_{\text{E}-\text{B}}|^2\gamma_{\text{E}}}\right)\geq\mathcal{R}\right\}\Pr\{\mathcal{B}_{\text{E}}\geq E_{\text{J}},Q_{\text{A}}>0\}\end{aligned}\qquad(8)$$

are performed offline and infrequently. The solution remains unchanged as long as the system's average parameters (e.g. average channel gains) remain unchanged.

The attacker can learn the average system parameters by monitoring the communications between Alice and Bob for a sufficiently long duration as explained next. By sensing the channel and knowing when the channel is busy and when it is not, the attacker can estimate $\lambda_A$ and the probability of the queue being empty/enonempty. Moreover, by overhearing the feedback channel from Bob to Alice, the link outage probability can be computed and the channel statistics can be estimated. The AWGN is due to thermal noise and its variance is fixed. Alice's transmit power can be estimated by measuring the signal power at Eve over different time slots. In summary, all the average system parameters can be estimated by the attacker. If one parameter is completely unpredictable, Eve can set this parameter to a known upper-bound value and operate under this assumption.

## IV. SIMULATION RESULTS AND CONCLUSIONS

We simulate our system using: $10^5$ channel realizations, $W=1$ MHz, $\kappa=1$ Watt, $\eta=0.6$, $E_{\text{d}}=5$ microjoules, $P_{\text{J}}/\kappa=10$ dB, and $P_{\text{A}}/\kappa=10$ dB. Each channel coefficient is assumed to be independent and identically distributed (i.i.d.) zero-mean circularly-symmetric complex Gaussian random variable with unit variance. To demonstrate the effectiveness of our proposed scheme in confusing the legitimate system without the need for external power, we set $E_{\text{const}}=0$, which means that the attacker does not get any energy from any external energy source.

In Fig. 1, we plot the secure throughput versus $\lambda_{\text{A}}$ for the proposed DoS/eavesdropping scheme under sensing and no sensing schemes. As shown in the figure, increasing $\lambda_{\text{A}}$, which is the amount of arrived data packets at Alice, increases the secure throughput of Alice. The performance of our proposed scheme with and without channel sensing is almost the same. Hence, we conclude that adding the sensing phase will not decrease the secure throughput of the legitimate system and, from a complexity viewpoint, Eve can skip this phase. The figure also shows the impact of increasing the transmit power at Alice. As shown in the figure, increasing the transmit power at Alice enhances the legitimate system's secure throughput at high data arrival rates. At low data arrival rates, using lower power at Alice enhances the security since the EH rate at Eve will decrease and, hence, the jamming probability decreases and the data decoding probability at Bob increases. On the other hand, at high data arrival rates, Alice's data queue will be saturated with data packets. Hence, the attacker increases his jamming probability which reduces the battery level at the attacker and reduces his ability to jam. Consequently, using higher power levels at Alice enhances the security due to the increase of the successful data decoding probability at Bob. However, since Alice does not know the used power-splitting ratio at Eve, she will not be able to compute the secure throughput. Hence, Alice will not be able to optimize her power level. We observe that the proposed scheme can efficiently degrade the secure throughput of the legitimate system. For example, when $\gamma_{\text{A}}=5$ dB, without attacks, the secure throughput of the legitimate system is $0.7$ packets/slot when $\lambda_{\text{A}}=0.7$ packets/slot, which implies that all the input data will be served. Under attacks, the secure throughput is $0.3$ packets/slot. Hence, the secure throughput loss is $57\%$, which demonstrates the efficiency and effectiveness of our proposed scheme.

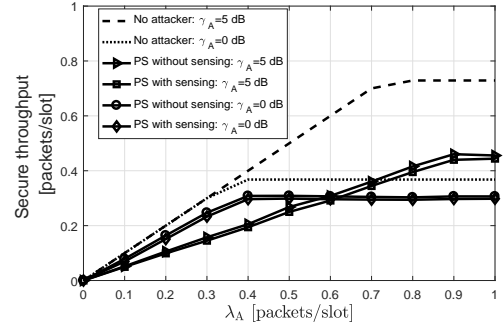

Fig. 1. Secure throughput versus $\lambda_{\text{A}}$. In the figure's legend, 'PS' refers to 'proposed scheme'.